\documentclass[journal,final]{IEEEtran}

\usepackage{graphicx}
\usepackage{subfigure}
\usepackage[tbtags]{amsmath}
\usepackage{amsfonts}
\usepackage{amssymb}
\usepackage{array}
\usepackage{algorithmic}
\usepackage{algorithm}
\usepackage{mathrsfs}
\usepackage{fancybox}
\usepackage{multirow}
\usepackage{cite}
\usepackage{color}
\usepackage{pifont}
\usepackage[pagewise]{lineno}
\usepackage{hhline}
\usepackage[implicit=false]{hyperref}
\usepackage[table]{xcolor}
\usepackage{hhline}
\usepackage{enumitem}
\usepackage{afterpage}
\usepackage{listings}
\usepackage{balance}

\begin{document}

%
\title{An embedded multichannel sound acquisition system for drone audition}

\author{  
Michael Clayton, Lin Wang, Andrew McPherson, Andrea Cavallaro        

\thanks{Manuscript received: December 25, 2020}
\thanks{The authors are with Centre for Intelligent Sensing, Queen Mary University of London, London, UK (e-mail: \{m.p.clayton, lin.wang, a.mcpherson, a.cavallaro\}@qmul.ac.uk)}
}

\maketitle

\begin{abstract}
Microphone array techniques can improve the acoustic sensing performance on drones, compared to the use of a single microphone. However, multichannel sound acquisition systems are not available in current commercial drone platforms. To encourage the research in drone audition, we present an embedded sound acquisition and recording system with eight microphones and a multichannel sound recorder mounted on a quadcopter. In addition to recording and storing locally the sound from multiple microphones simultaneously, the embedded system can connect wirelessly to a remote terminal to transfer audio files for further processing. This will be the first stage towards creating a fully embedded solution for drone audition. We present experimental results obtained by state-of-the-art drone audition algorithms applied to the sound recorded by the embedded system.       
\end{abstract}


\section{Introduction} \label{sec:intro}
The use of drones for remote sensing has substantially increased in the past decade, with operation in broadcasting, surveillance, inspection, and search and rescue~\cite{floreano2015science}. Sensing is primarily based on cameras (optical and thermal) and lasers\cite{Parascandolo16, li2017visual, misra2018aerial, wang2018tracking, sanchez2017multi}, whereas microphones are rarely used because of  the inherently challenging sound sensing conditions~\cite{wang2018acoustic}. When visual data is unreliable due to low light, poor weather conditions or visual obstructions~\cite{deleforge2019audio}, drone audition would greatly benefit the above-mentioned applications.  One of the main obstacles when capturing audio on a drone is the strong ego-noise created by the rotating motors, propellers and the airflow during flight. The ego-noise masks the target sound sources and causes poor recording quality. 

Microphone array techniques can be used to improve the drone audition performance through sound enhancement \cite{schmidt2018novel, kang2019software, hioka2019design, wang2020deep, wang2020blind, wang2017microphone} and sound source localization \cite{yen2020noise, wang2017time, strauss2018dregon,  wakabayashi2020drone, furukawa2013noise}. An important bottleneck for deploying microphone array algorithms on drones is the requirement of a multichannel sound acquisition system to enable sampling the sound from multiple microphones simultaneously and convert it to multichannel digital signals before further processing. The sound acquisition system needs to fly with the drone, which imposes additional constraints on the size and weight of the system. To the best of our knowledge, there is  no dedicated multichannel sound acquisition device available in current commercial drone platforms. Researchers have to design and implement their own hardware systems for data collection on drones, and the processing of the data is often done offline after the flight due to limited computational resources onboard.

To conduct and encourage the research in the field of drone audition, we designed an embedded multichannel sound acquisition system that is suitable for drone audition and can be mounted on a drone for acoustic sensing during flight. The system is designed based on Bela \cite{mcpherson2015environment}, an embedded computing platform dedicated to audio processing, and can accommodate up to eight microphones placed in arbitrary shapes. The system can record and store the sound {\em locally} for on-device processing; and can also transfer the recorded sound file via wireless communication to a remote terminal. In the remainder of the paper, we disclose the technical details for hardware, software design and development.

The paper is organized as follows. Sec.~\ref{sec:related} reviews related works. Sec.~\ref{sec:hardware} and Sec.~\ref{sec:software} present the hardware and software design of the embedded system. Sec.~\ref{sec:experiment} presents real data collection with the hardware and presents baseline processing results with state-of-the-art drone audition algorithms.

\begin{table*}[t]
\centering
\caption{Existing multichannel sound acquisition systems on drones. Q - Quadcoptor; H - Hexacopters; O - Octocopter}
\label{table:1}

\begin{tabular}{|c|c|c|c|c|c|c|}
\hline
{\bf Ref} & {\bf Number of} & {\bf Shape } & {\bf  Placement} & {\bf Audio interface} & {\bf Drone Type} & {\bf Remark} \\
& {\bf microphones} & {\bf of the array} & {\bf  of the array} &  &  & \\
\hline
\cite{hioka2019design} & 6 & T-shape & Side  & Zoom H6 & Self-assembled (Q) & Portable recorder\\ 
\cite{wang2018acoustic} & 8 & Circular & Top & Zoom R24 & 3DR Iris (Q) & Portable recorder\\ 
\cite{andra2019feasibility} & 6 & Circular (fixed) & Top & ReSpeaker + Raspberry Pi &  Self-assembled (O) & Intelligent voice interface \\ 
\cite{tan2019efficient} & 7 & Circular (fixed) & Side  & UMA-8 Mic array + Raspberry Pi &  Self-assembled (Q) & Intelligent voice interface \\ 
\cite{salvati2019acoustic} & 8 & Circular & Below  & MiniDSP USBStreamer I2S-to-USB & Matrice 100 (Q) & Sound card \\ 
\cite{strauss2018dregon} & 8 & Cubic & Below  & 8SoundsUSB & MK-Quadro (Q) & Sound card \\
\cite{ruiz2018aira} & 8 & Circular & Top, below, side  & 8SoundsUSB & Matrice 100 (Q) & Sound card \\ 
\cite{hoshiba2017design} & 12 & Spherical & Side  & RASP-ZX & Surveyor MS-06LA (H) & Sound card \\ 
\cite{okutani2012outdoor} & 8 & Circular & Side  & RASP-24 & Parrot AR Drone (Q) & Sound card\\ 
\cite{wakabayashi2020drone} & 16 & Octagon & Side  & RASP-ZX & Surveyor MS-06LA (H) & Sound card \\
\hline
{\bf Proposed} & 8 & Circular & Top  & Bela & Matrice 100 (Q) & Sound card \\
\hline
\end{tabular}
\end{table*}

\section{Related work} \label{sec:related}
Three types of audio hardware are employed: multichannel sound card, intelligent multichannel voice interface, and portable multichannel sound recorder.   

\subsubsection{A portable multichannel sound recorder} This is the easiest way to capture sound from drones as there is no requirement for any configuration of the system, e.g. Zoom H6~\cite{hioka2019design} and Zoom R24~\cite{wang2018acoustic, wang2019audio}. The hardware supports arbitrary array topology. The drawback is the the hardware can only achieve recording and does not support sound processing. Another drawback is that the hardware, e.g. Zoom R24, is usually too heavy a payload for the drone to fly.

\subsubsection{Intelligent multichannel voice interface} This type of hardware integrates the microphone array and sound processing into a compact IC board, e.g. ReSpeaker~\cite{andra2019feasibility} and UAM-8~\cite{tan2019efficient}. This hardware usually requires an additional controller, e.g. Rasberry Pi, for sound acquisition and sound processing. This hardware is also usually easy to use and configure for audio purposes. One of the main advantages is that the hardware is also very compact and light-weight, and is suitable to fly with the drone. The drawback is the topology of the array is fixed, which limits the performance and flexibility of microphone array algorithms.   

\subsubsection{Multichannel sound card} This is the most popular approach for sound recording on drones, using e.g. RASP series~\cite{hoshiba2017design, okutani2012outdoor, wakabayashi2020drone}, 8SoundUSB~\cite{ strauss2018dregon, ruiz2018aira}, USB Streamer~\cite{ salvati2019acoustic}. This hardware supports arbitrary array topology along with sound acquisition and sound processing. The main drawback is the user requires knowledge of the hardware circuit design. This particular hardware also requires an operating system to control sound recording and processing, e.g. the RASP series is used in combination with the HARK system~\cite{nakadai2017development}. A good understanding of the back-end driver is necessary. The lack of related resources is also intimidating for algorithm designers.

\section{Hardware design} \label{sec:hardware}
\begin{figure}[tb]
\vspace{0.2in}
\centering
\includegraphics[trim=0cm 0cm 0cm 0cm, clip=true, width=0.48\textwidth]{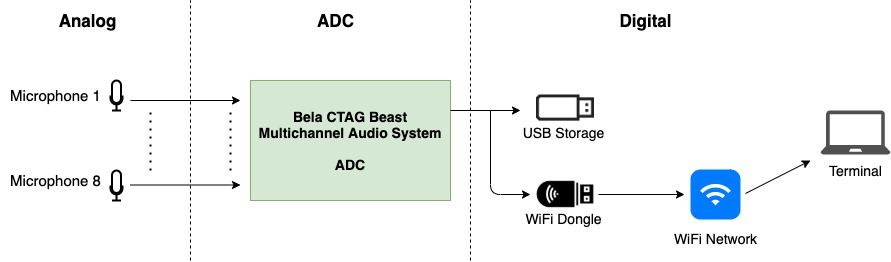} \\
\caption{\small{Architecture of the multichannel sound acquisition system.}}
\label{fig:architecture}
\end{figure}

\begin{figure}[tb]
\vspace{0.2in}
\includegraphics[trim=5cm 1.2cm 12cm 4cm, clip=true, width=0.48\textwidth]{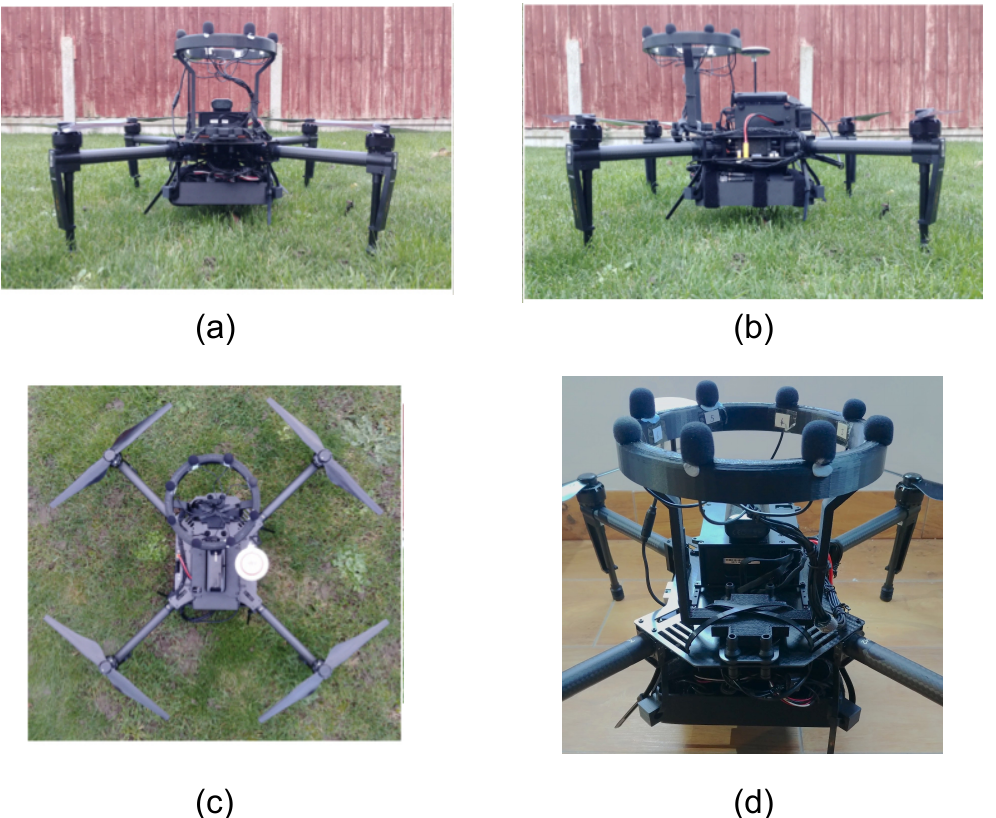}
\caption{Real objects of the multichannel sound acquisition system. (a) Front view; (b) Side view; (c) Top view; (d) Microphone array.}
\label{fig:object}
\end{figure}

\begin{figure}[tb]
    \centering
     \includegraphics[width=0.45\textwidth]{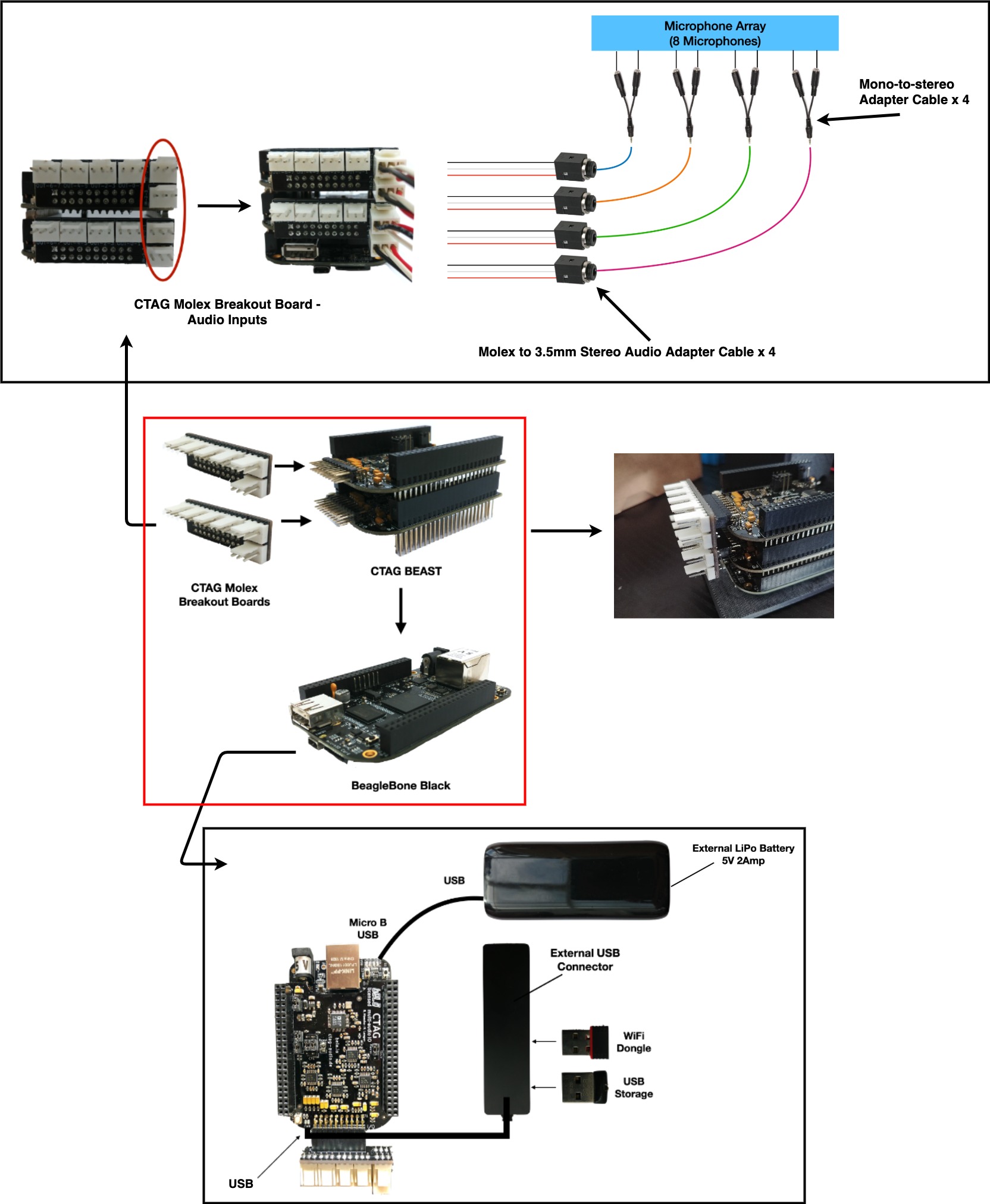}
    \caption{\small{Bela multichannel audio hardware system assembly and peripheral. The core processing part is highlighted in the red box. }}
    \label{fig:connection}
\end{figure}

Fig.~\ref{fig:architecture}  and Fig.~\ref{fig:object} illustrate the architecture and the real objects of the multichannel sound acquisition system, respectively. The system mainly consists of three parts: the microphone array, the drone, the hardware tray containing the Bela sound acquisition system and the cables. Table~\ref{tab:component} lists the components used by the system. Fig.~\ref{fig:connection} illustrates the Bela hardware system assembly and peripheral connections. 

\begin{table}[tb]
\centering
\caption{Components used in the hardware system.}
\label{tab:component}

\begin{tabular}{|m{2.5cm}|m{2.5cm}|m{2.5cm}|}
\hline
Component & Type & Functionality  \\
\hline
Drone  & Matrice 100 & /  \\
\hline
Microphones (8) & Lapel microphones & /  \\
\hline
Array frame & 3D printing & Holding microphones  \\
\hline
Hardware tray & 3D printing & Holding hardware and  cables  \\
\hline
Bela & BeagleBone Black & 1GHz ARM Cortex-A8 processor  \\
\hline
CTAG Beast (2) & / & Multichannel audio acquisition  \\
\hline
CTAG Molex breakout board (2) & / & Audio inputs  \\
\hline
Molex to 3.5mm adapter cable (4) & / & Connects microphones to Bela  \\
\hline
Mono to stereo adapter cable (4) & / & Split stereo signal to mono signal  \\
\hline
USB LiPo battery & 5V, 2Amp & Provide power to Bela CTAG BEAST  \\
\hline
USB storage & / & Save and store recorded audio files locally  \\
\hline
WiFi Dongle & / & Wireless connection to bela IDE  \\
\hline
\end{tabular}
\end{table}

\subsection{Microphone array and drone}
We use a circular microphone array consisting of eight Boya BY-M1 lapel microphones that are each powered by an LR44 (1.5V) battery. A balanced audio signal is provided from the microphones. The diameter of the array is 16.5 cm. The microphone array frame is 3D printed and constructed from Acrylonitrile butadiene styrene (ABS). The array is mounted on top of the drone to avoid the air flow from the rotating propellers blowing downward~\cite{wang2016ear}. The vertical distance from the array to the drone body is 18 cm. For the drone, we use DJI Matrice 100, which has a payload capacity of 1 kg.  

\subsection{Bela-based sound acquisition system}

The sound acquisition system consists of four units: the core processing unit,  the storage and transmission unit, and the hardware tray. Fig.~\ref{fig:connection} illustrates the hardware connections. 

\subsubsection{Core processing unit}
The core processing unit consists of one Beaglebone device flashed with the latest Bela software. To access multichannel audio, Bela uses a customized expansion board called CTAG BEAST, featuring an audio codec with 4 audio input and 8 audio output channels. One CTAG BEAST consists of 2 x CTAG FACE capes pre-configured for use as a BEAST \cite{langer2017ctag}, two CTAG Molex breakout boards, and one external LiPo power battery. 

Bela\footnote{Bela is an embedded audio programming and processing platform invented by academia from Queen Mary University of London\cite{mcpherson2015environment}. The compact size, light-weight, low-latency and multichannel sound acquisition makes it suitable for sound processing on drones\cite{mcpherson2016action}. Bela also comes with a user friendly browser-based Integrated Development Environment (IDE), which is used for easy access for editing, building and managing the system.  For this reason, we decided to develop a multichannel sound acquisition system based on the Bela device. This is the first time the Bela system has been applied to robotic platforms assisting audition.} is a dedicated audio processing platform based on BeagleBone Black (BBB) single-board computer, which is featured by a 1GHz ARM Cortex-A8 processor, two PRUs (Programmable Realtime Units), 512 MB RAM, and a diverse range of on-board peripherals. Bela is used for controlling the sound acquisition and audio processing. Bela is externally powered by a LiPo USB battery that operates at 5V and 2Amp for stability and powering the USB peripherals. The Bela configuration only requires 5V / 300 - 400mA of power for operation.
 
The audio codec operates at 48 kHz sampling rate with 16 bits analogue-to-digital converter (ADC) and digital-to-analogue converter (DAC) conversion. To accommodate 8 microphone inputs, two CTAG FACE capes are stacked on top of each other and connected with Bela via the onboard metal contacts.

\subsubsection{Storage and wireless unit}
A external USB hub is connected to the USB socket of the Beaglebone device. The hub accommodates a USB storage stick, which stores the recording locally, and a USB WiFi Dongle, eliminating the need for a hard-wired connection to the system IDE and enabling the recorded audio to be transferred to a remote processing terminal. 

\subsubsection{Hardware tray}
A hardware tray is designed to accommodate the Bela system and the cables. The tray contains a Bela enclosure (made from ABS) and shock case (made from Thermoplastic Polyurethane - TPU) to aid with protecting the hardware from impacts in the event of a crash. The tray is produced with 3D printing. 

\section{software design} \label{sec:software}

The software design has three objectives: to run the code in a stand-alone device; to record the sound locally to the USB storage; to transfer the sound via WiFi to a remote terminal. All the objectives are achieved with the assistance of the Bela Integrated Development Environment (IDE). 

\lstset{
  basicstyle=\footnotesize,
  breaklines=true
}

\subsection{IDE for stand-alone processing}
\label{sec:ide}

 \begin{figure}[tb]
    \centering
    \includegraphics[trim=0cm 0cm 0cm 0cm, clip=true, width=0.48\textwidth]{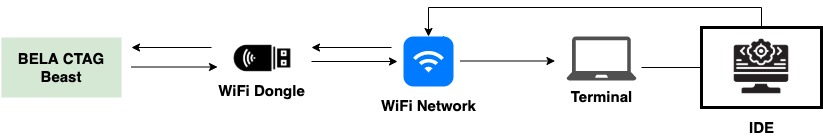}
    \caption{\small{Interacting with Bela from a local computer via wireless network.}}
    \label{fig:Wifisigflow}
\end{figure}

\begin{figure}[tb]
    \centering
    \includegraphics[width=0.48\textwidth]{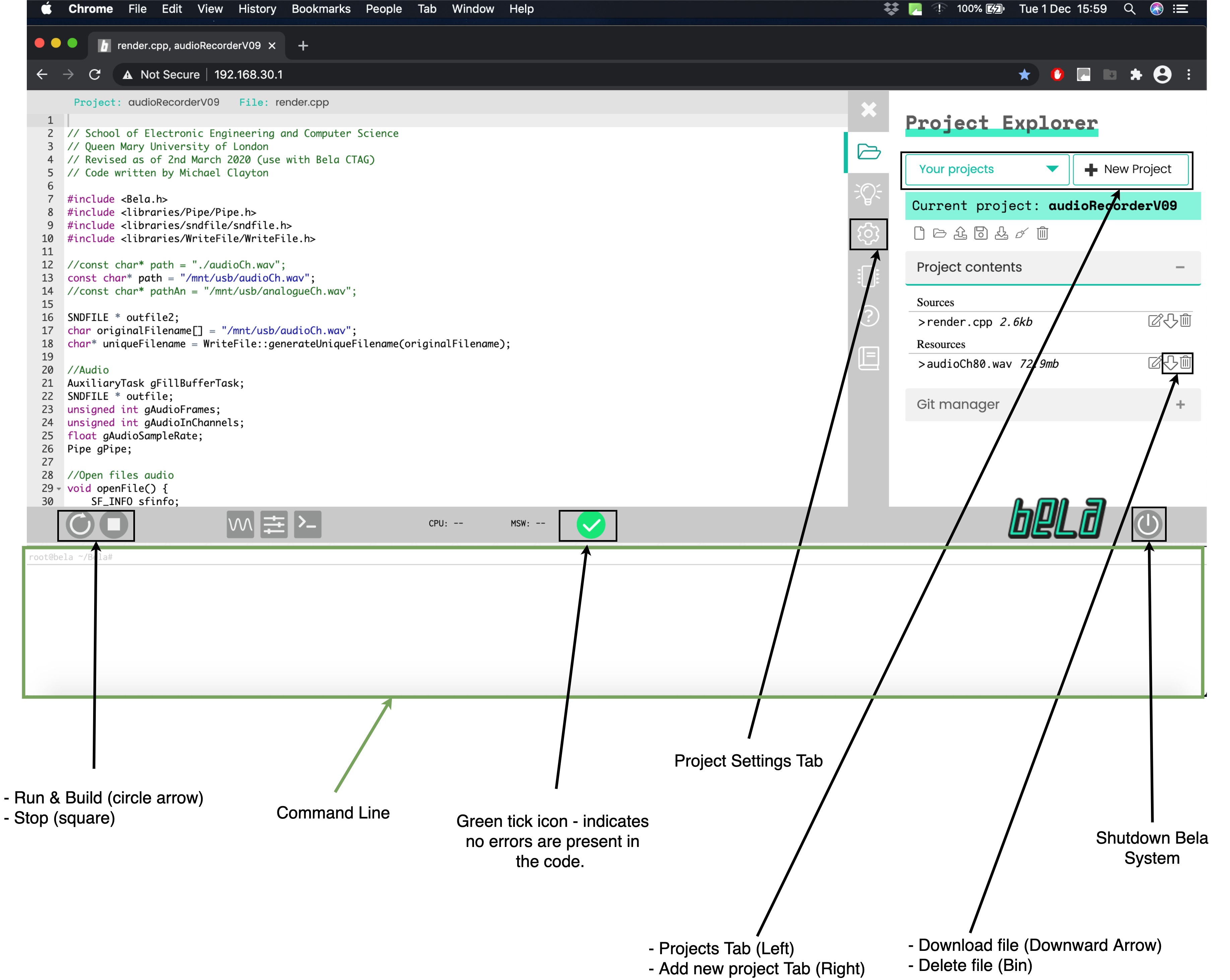}
    \caption{\small{Bela Integrated Development Environment.}}
    \label{fig:Bela_IDE}
\end{figure}

The Bela IDE (Fig.\ref{fig:Bela_IDE}) is a browser based integrated development environment with features that allow for editing, building and managing projects easily from a ground station (remote terminal) via a self-organized wireless network. 

The IDE software is pre-installed at the Bela device, along with an operation system. Following the steps in the tutorial\footnote{\href{www.eecs.qmul.ac.uk/~linwang/download/bela_documentation.pdf}{www.eecs.qmul.ac.uk/$\thicksim$linwang/download/bela\_documentation.pdf}}, we set up a self-organized wireless network through a WiFi dongle mounted on the Bela device. Upon system boot, Bela starts a NodeJS server that allows  connection to its system from a ground station via the wireless network. The WiFi is setup as a peer-to-peer connection to ensure that the board acts as a dynamic host configuration protocol (DHCP) server.  
 
To connect to the Bela device from the ground station, we first need to select the WiFi network hosted by the Bela system. After connection, the IDE can be loaded by entering the IP address of the host device from the web browser. The IDE interface (Fig.~\ref{fig:Bela_IDE}) will appear automatically at the web browser of the ground station. 
 
After compiling, building and running a project from the IDE. The project can be set to run on boot in the IDE Settings tab by selecting the desired project in the drop down menu. The program will operate on Bela without connecting to the ground station as long as the external power is provided.
 
\subsection{Sound recording}

The code that enables the Bela to function as a multichannel recording device is written in C++. This allows for quick access in the event that the system requires any modifications, creating a flexible system. The source code for sound recording is given in the Appendix, with the processing flow shown in  Fig.~\ref{fig:recording_flow}. In brief, after importing the required library, configuring global variables and file path, the program sets up the recording task to capture the multichannel audio data, writes the stream to the audio buffer (memory block), and stores the data in the pre-defined file path. Once the recording is finished, a clean-up function finalizes the writing process and closes the file. 

The IDE enables the user to start/stop recording, change settings and download audio files directly from the system among some other features. After building and running the project, the recording will start by writing the digital audio to the specified system path. Pressing the stop icon in the IDE will stop the recording process. The audio data is continuously written to the local storage during recording. Once the stop button is pressed the .wav file is finalised and closed. 

\begin{figure}[tb]
    \centering
    \includegraphics[trim=0cm 0cm 0cm 4.8cm, clip=true, width=0.45\textwidth]{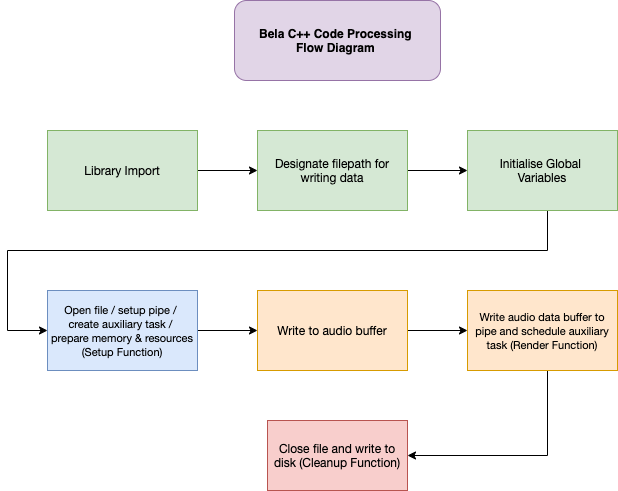}
    \caption{\small{Processing flow for sound recording with Bela.}}
    \label{fig:recording_flow}
\end{figure}

\subsection{WiFi Network Connection}

The WiFi connection enables the user to access the Bela system through the IDE without a hard-wired connection. Instead of recording the audio to the USB storage, we can alter the target file-path to the default RAM memory of the Bela device (see Appendix) in order to have the file appear and update during the recording process within the resources section of the project explorer tab in the IDE. The network connection is continuous to allow the user to change different functions in the IDE.

The current WiFi signal is able to achieve an operational range of 20 metres between ground station and the drone. When the network connection is lost momentarily, the IDE user interface stops updating the user about the project running and dependent on filepath selection, the file size of the current recording. The IDE recovers after coming back into WiFi signal range.  When the wireless network is re-established the existing file has essentially continued running on the system and the IDE user interface resumes updating the recording progress of the file. The WiFi connection is not required to conduct the recording itself but to monitor its progress.

\section{Experiment} \label{sec:experiment}

\subsection{Setup}
\begin{figure}[tb]
    \centering
    \includegraphics[trim=0cm 3cm 0cm 3cm, clip=true, width=0.4\textwidth]{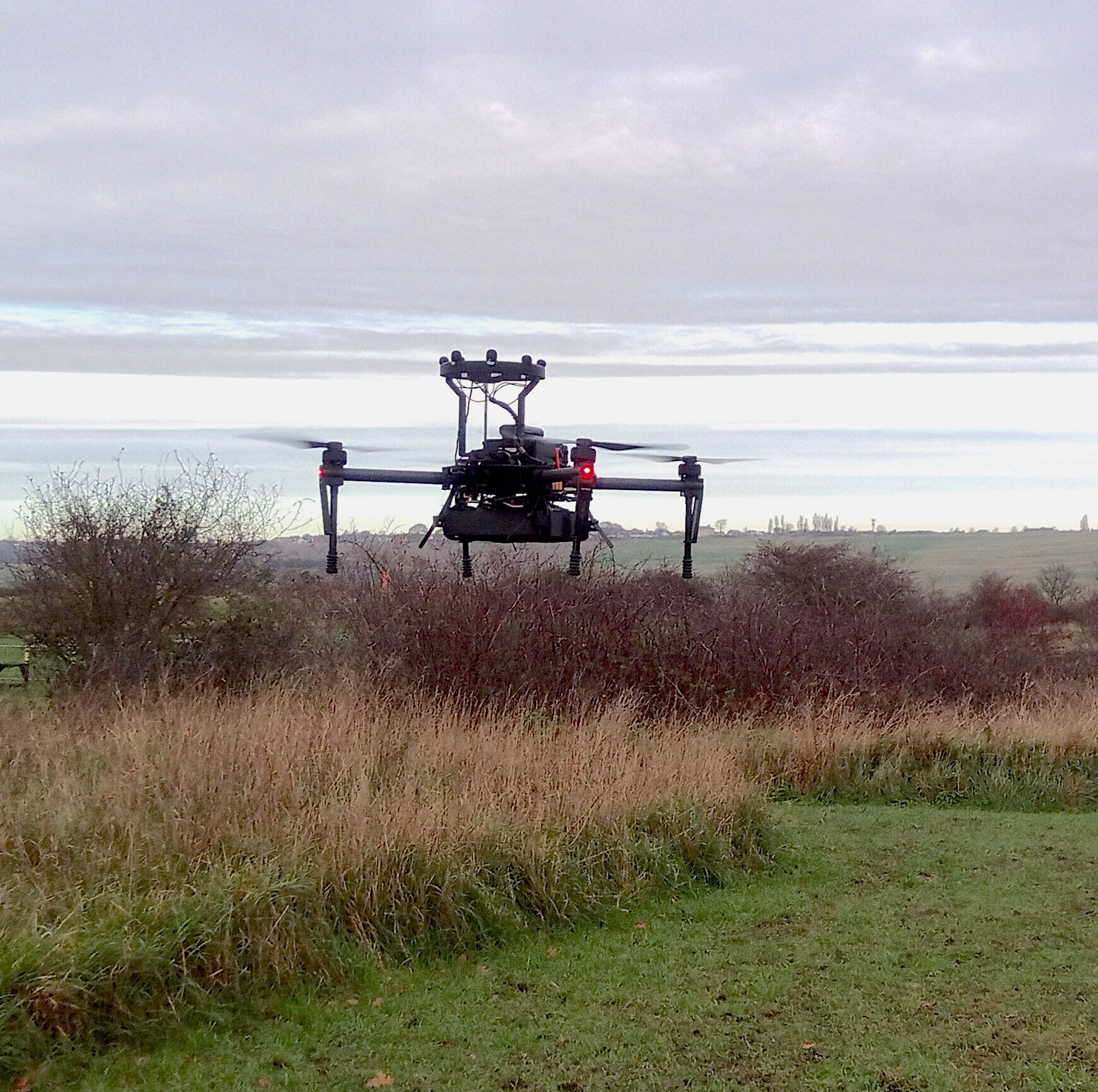}
    \caption{\small{A drone with microphone array hovering in the air during recording.}}
    \label{fig:drone_hovering}
\end{figure}

To verify the validity of the developed hardware system, we conduct in-flight testing and recording. We record the ego-noise and the speech, separately. When recording the ego-noise, the altitude of the drone during flight is maintained at about 2 meters above the ground via the flight controller (Fig.~\ref{fig:drone_hovering}). We record two types of ego-noise: drone hovering and drone moving. In the former case, the drone is hovering in the air using the GPS stabilised mode with additional manual input (correcting small drift) to allow the drone to remain reasonably stable throughout the recording. In the latter case, the drone is moving in the air at a speed of around 1 meter/second, with random rotation and tilting during flight. When recording the speech-only data, the drone is muted on the ground and a loudspeaker plays sound at a distance of 2 meters. The original sampling rate is 48 kHz. The audio is downsampled to 8 kHz before processing. All the analysis is completed offline and not on the Bela system. The sound recording is available online\footnote{\href{www.eecs.qmul.ac.uk/~linwang/bela.html}{www.eecs.qmul.ac.uk/$\thicksim$linwang/bela.html}}.   

\subsection{Ego-noise analysis}
\begin{figure}[tb]
    \centering
    \includegraphics[trim=2.4cm 6cm 3cm 5.9cm, clip=true, width=0.48\textwidth]{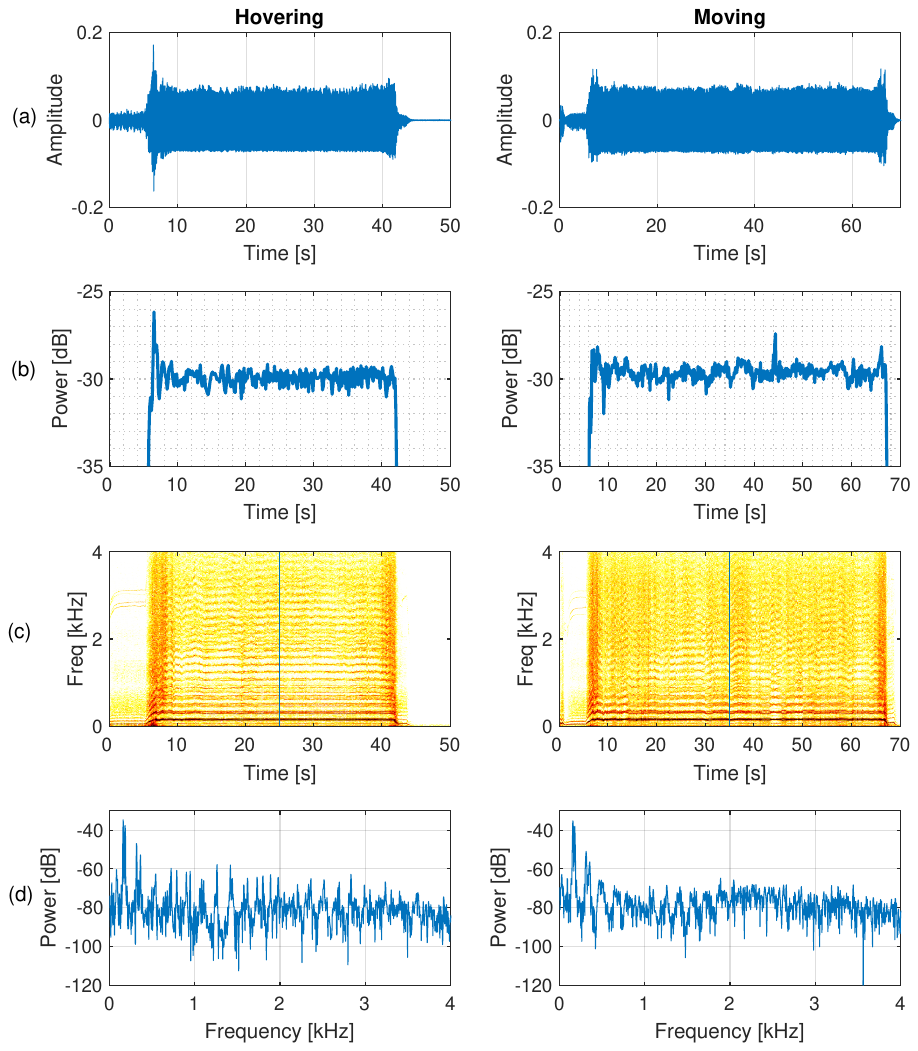}
    \caption{Visualization of the ego-noise when the drone is hovering and moving. (a) Time-domain waveform; (b) Power plot; (c) Time-frequency spectrogram; (d) Frequency-domain plot.}
    \label{fig:spec}
\end{figure}

Fig.~\ref{fig:spec}(a) depicts time-domain waveform of the ego-noise recorded at the hovering and the moving status. Fig.~\ref{fig:spec}(c) plots the time-frequency domain spectrogram, which is computed with a moving window of 128 ms and half overlap. Fig.~\ref{fig:spec}(b) plots the power at each time frame. Fig.~\ref{fig:spec}(d) plots the frequency-domain spectrum at the 25-th second and 35-th second of the two ego-noises, respectively. 

From Fig.~\ref{fig:spec}(b), the power of the ego-noise does not show big differences at the hovering status and the moving status. The mean and standard deviation of the power across time frames (10$s$-40$s$) are -29.9 dB and 0.41 dB, respectively, at the hovering status. The mean and standard deviation of the power across time frames (10$s$-60$s$) are -29.6 dB and 0.42 dB, respectively, at the moving status. For the moving status, we observe a sudden rise of the power at 45$s$, which is possibly due to the rotation operation of the drone. 

From Fig.~\ref{fig:spec}(c), it can be observed that the ego-noise consists of multiple harmonics. Since the four motors might operate at a slight different rotating speed, the harmonic ego-noise presents several pitches, which can be verified from Fig.~\ref{fig:spec}(d). At the hovering status, the pitch of the ego-noise remains stable. At the moving status, the pitch of the ego-noise varies with time, depending on flight status of the drone. 

The recording is made in an outdoor environment with a light breeze present. However, from the spectrogram of the recording we do not observe an evident influence of the wind in the low frequency. This is possibly due to the windshield worn by each microphone and also the placement of the microphones on top of the drone.

\begin{figure}[tb]
    \centering
    \includegraphics[trim=5cm 10cm 5cm 11cm, clip=true, width=0.4\textwidth]{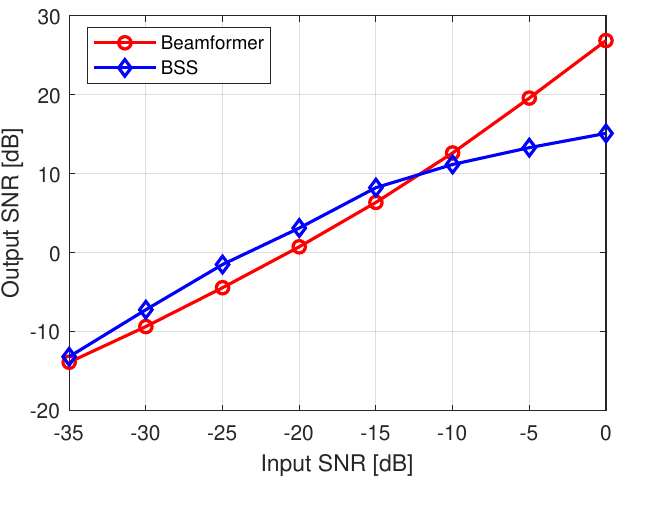}
    \caption{Benchmark performance achieved by two spatial filters at various input SNRs. }
    \label{fig:perf}
\end{figure}

\begin{figure}[tb]
    \centering
    \includegraphics[trim=4.5cm 9cm 5.0cm 9cm, clip=true, width=0.48\textwidth]{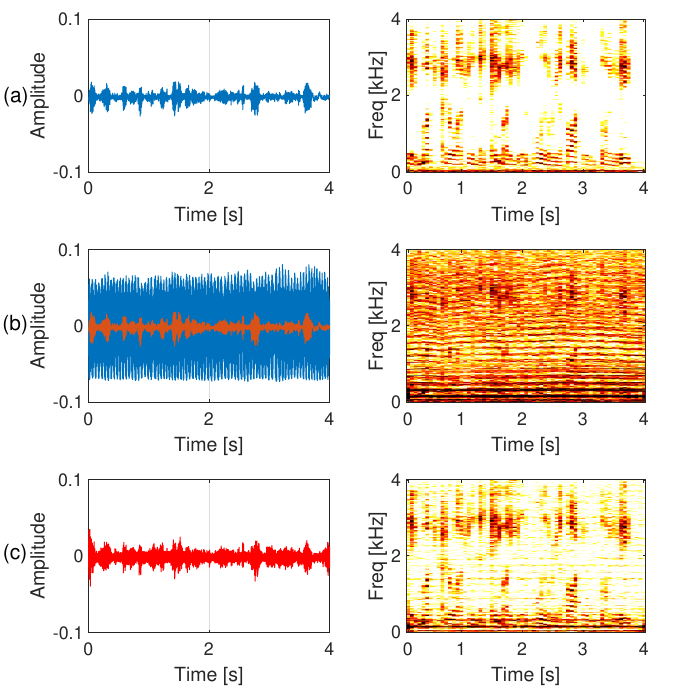}
    \caption{Processing results (beamformer) for input SNR -20 dB. The output SNR is 3.1 dB. (a) Clean speech; (b) Noisy signal before procesisng; (c) Noisy signal after processing. }
    \label{fig:result}
\end{figure}

\subsection{Processing results}
We synthesize a noisy signal at the microphones by adding the ego-noise (hovering status) and the speech at different input SNRs, which vary from -35 dB to 0 dB, with an interval of 5 dB. The testing signal is 25 seconds long. We employ a block-wise processing strategy, using a non-overlap sliding block of 4 seconds. For simplicity, we just verify the performance of two benchmark spatial filters enhancing the target sound from the ego-noise. The first spatial filter is a beamformer based on Multichannel Wiener filtering~\cite{wang2017microphone} which computes the correlation matrix of the target sound and the noise separately assuming the speech-only and noise-only signals are available. The second spatial filter is based on blind source separation (BSS)~\cite{wang2020blind}, assuming the permutation ambiguities can be perfectly solved by referencing to the speech-only signals. The speech enhancement performance is evaluated with the SNR measure, which is defined, given speech $s(n)$ and noise $v(n)$, as~\cite{wang2015noise}
\begin{equation}
    \text{SNR} = 10\log_{10}\frac{\sum_{n}{s^2(n)}}{\sum_{n}{v^2(n)}}
\end{equation}
We average the output SNR across all the processing blocks. 

Fig.~\ref{fig:perf} depicts the output SNR achieved by the two spatial filters at different input SNRs. BSS performs slightly better than beamformer when the input SNR is lower than \text{-15} dB, while beamformer performs better at higher input SNRs. On average, the two spatial filters improve the SNR by about 20~dB. 

Fig.~\ref{fig:result} illustrates exemplar processing results at input SNR \text{-20} dB with the beamformer. The output SNR is 3.1 dB. Fig.~\ref{fig:result}(b) shows that the speech signal is completely buried in the ego-noise in the time-domain waveform and not distinguishable between each other in the time-frequency spectrogram. Fig.~\ref{fig:result} shows the enhanced speech after processing, where the speech is better observed in the time-frequency spectrogram. 

It should be noted that the two spatial filters are estimated with ideal assumption (i.e. the correlation matrix of the target and the noise are known) and thus set the benchmark of the performance of spatial filtering. In practice, the correlation matrices of the target and noise have to be estimated from the noisy data, which leads to performance drop in low-SNR scenarios~\cite{wang2017microphone}. A comprehensive evaluation will be left for future work. 

\section{Conclusion} \label{sec:conclusion}

We present an embedded multichannel sound acquisition system that can fly with the drone. The system can accommodate up to 8 microphones placed in an arbitrary shape, record the sound locally and can transfer the recorded file to a remote terminal via a self-organized wireless network. Experimental results with recordings made with this hardware verify its validity. This will be the first stage towards creating a fully embedded solution for drone audition. 

Future work would be to conduct a comprehensive evaluation of the state-of-the-art algorithms for ego-noise reduction and to optimize the code for real-time processing at Bela, which is able to process audio at very low latency ($<$1 millisecond)~\cite{mcpherson2015environment}. The size and weight of the system can be furthered reduced by designing the microphone array circuit manually. 

\section*{Appendix: Sound recording}
\begin{figure}[tb]
    \centering
    \fbox{
    \includegraphics[trim=1cm 1cm 0.5cm 0.5cm, clip=true, width=0.48\textwidth]{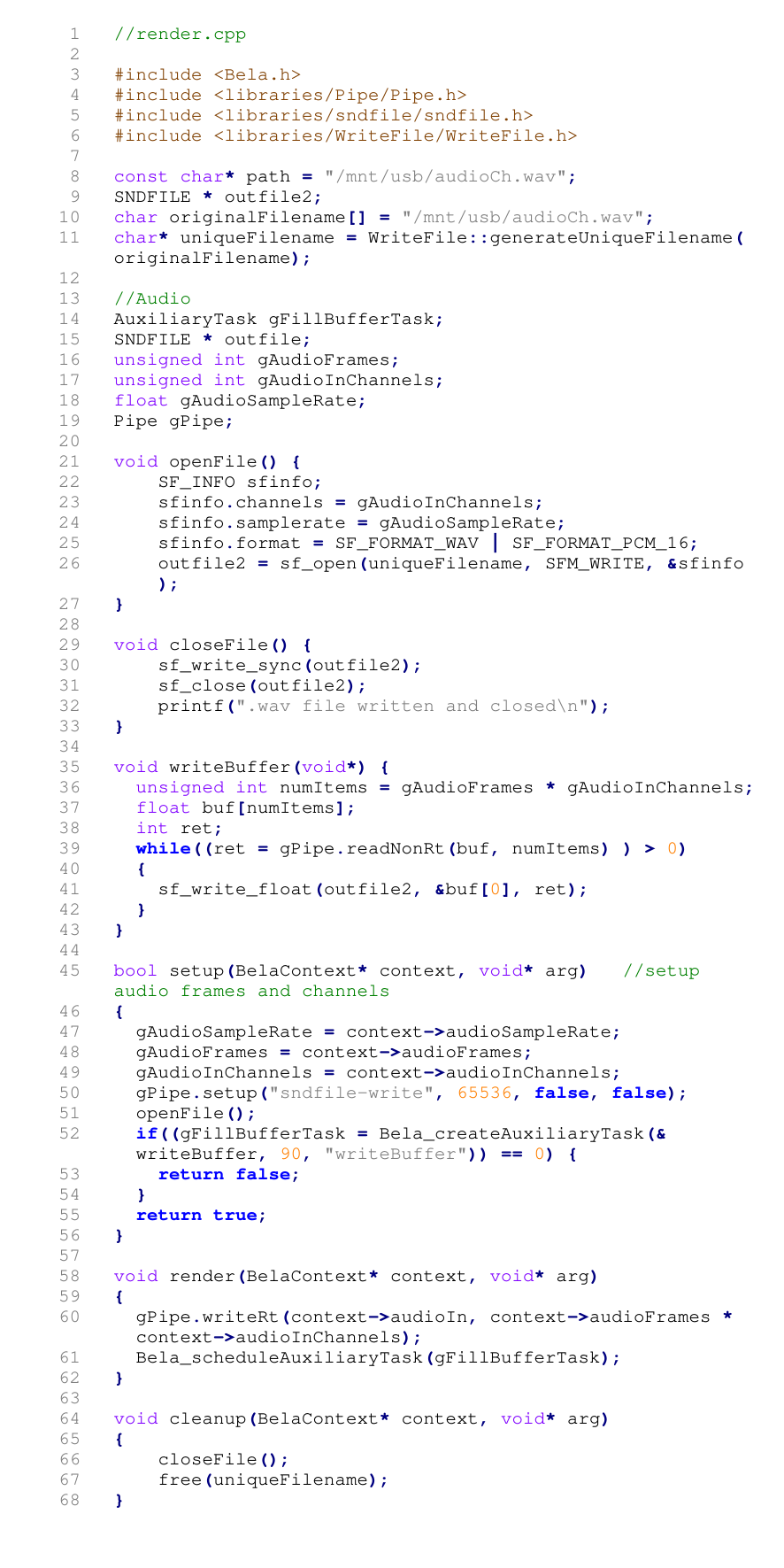}}
    \caption{\small{Source code in \texttt{render.cpp}}}
    \label{fig:render_cpp}
\end{figure}

\lstset{
  basicstyle=\scriptsize,
  breaklines=true
}

Fig.~\ref{fig:render_cpp} lists the C++ source code file that is used for multichannel sound recording with the Bela device. There are several crucial configurations for sound recording: the file path, the number of channels and the sampling rate. The file path can be configured by setting the global variable {\ttfamily{const char* path}} in the source code, e.g. the command \texttt{const char* path = "/mnt/usb/audiofilename"} sets USB storage as the file path. The system can automatically recognize the the amount of active audio inputs and thus does not need to configure the number of channels. The sampling rate can be configured by setting the global variable {\ttfamily{gAudioSampleRate}}. The ADC and DAC gain is adjustable within the IDE settings tab. Once the recording is finished, the file on the USB storage can be downloaded through IDE after copying them to the project folder, e.g. using the command \texttt{cp /mnt/usb/audiofile.wav /root/Bela/projects/projectname/}. Alternatively, we can remove the USB storage from Bela and insert it into a computer for data transfer. 

A breakdown and interpretation of the source code is given below.  

\vspace{0.2in}
\begin{lstlisting}[frame=single, language=C++]
#include <Bela.h>
#include <libraries/Pipe/Pipe.h>
#include <libraries/sndfile/sndfile.h>
#include <libraries/WriteFile/WriteFile.h>
\end{lstlisting}

There are four main library imports that are used in the code. Bela.h is the central control code for hard real-time audio on BeagleBone Black using PRU and Xenomai Linux extensions.
The pipe library enables the use of a bi-directional pipe that allows for data to be exchanged between realtime and non-realtime thread. The Writefile library is imported to enable the use of the generateUniqueFilename function, that returns a unique filename appending a number at the end of the original filename. This is important in order to avoid overwriting existing recordings. The sndfile library allows the use of the libsndfile API (application programming interface) which is designed to allow the reading and writing of many different sampled sound file formats. 

\vspace{0.2in}
\begin{lstlisting}[frame=single, language=C++]
const char* path = "/mnt/usb/audioCh.wav";
SNDFILE * outfile2;
char originalFilename[] = "/mnt/usb/audioCh.wav";
char* uniqueFilename = WriteFile::generateUniqueFilename(originalFilename);
\end{lstlisting}

The file path is created first, in this case the usb is used as the main storage. The outfile2 is the name of the reference for the SNDFILE pointer. The path to the original filename is assigned and then the generateUniqueFilename function is called on the original filename to return a unique filename. 

Instead of recording the audio to the USB storage, the filepath can be altered to \texttt{const char* path = "./audiofilename"} in order to have the file appear and update during the recording process within the resources section of the project explorer tab in the IDE.

\vspace{0.2in}
\begin{lstlisting}[frame=single, language=C++]
AuxiliaryTask gFillBufferTask;
unsigned int gAudioFrames;
unsigned int gAudioInChannels;
float gAudioSampleRate;
Pipe gPipe;
\end{lstlisting}

The global variables are established for use later in the code. The gFillBufferTask is a Auxiliary task variable that is used to write to the audio buffer. The gPipe, gAudioSampleRate, gAudioInChannels and gAudioFrames are global variables that utilise the established behaviours present in their corresponding library class files.

\vspace{0.2in}
\begin{lstlisting}[frame=single, language=C++]
void openFile() {
    SF_INFO sfinfo;
    sfinfo.channels = gAudioInChannels;
    sfinfo.samplerate = gAudioSampleRate;
    sfinfo.format = SF_FORMAT_WAV | SF_FORMAT_PCM_16;
    outfile2 = sf_open(uniqueFilename, SFM_WRITE, &sfinfo);
}
\end{lstlisting}

The openFile function details the \lstinline{SF_INFO} structure and the specified file format, sample rate, amount of channels. The \lstinline{sf_open} function opens the sound file at the specified path and utilises the write only mode \lstinline{SFM_WRITE} and the sfinfo structure for passing data between the calling function and the library when opening the file for in this case writing. 

\vspace{0.2in}
\begin{lstlisting}[frame=single, language=C++]
void closeFile() {
    sf_write_sync(outfile2);
    sf_close(outfile2);
    printf(".wav file written and closed\n");
}
\end{lstlisting}

The closeFile function closes and writes the file to disk. \lstinline{sf_write_sync} allows for the file if it is opened using \lstinline{SFM_WRITE} to call the operating system's function to force the writing of all file cache buffers to disk. \lstinline{sf_close} closes the file, deallocates its internal buffers and returns 0 on success or an error value otherwise. 

\vspace{0.2in}
\begin{lstlisting}[frame=single, language=C++]
void writeBuffer(void*) {
  unsigned int numItems = gAudioFrames * gAudioInChannels;
  float buf[numItems];
  int ret;
  while((ret = gPipe.readNonRt(buf, numItems) ) > 0)
  {
    sf_write_float(outfile2, &buf[0], ret);
  }
}
\end{lstlisting}

The writeBuffer function essentially writes data to the buffer. This function calculates the number of items by multiplying the audio frames and audio input channels. A buffer array holds the audio frame and input channel items. An integer variable is declared with the purpose of holding the number of items. The while loop will loop through the block of code as long as the specified condition is true. The readNonRt reads data from the non-realtime side. The \lstinline{sf_write_float} function writes the data in the array pointing to the pointer of the file. For items-count functions, the items parameter specifies the size of the array and must be an integer product of the number of channels or an error will occur. 

\vspace{0.2in}
\begin{lstlisting}[frame=single, language=C++]
bool setup(BelaContext* context, void* arg)
{
 
  gAudioSampleRate = context->audioSampleRate;
  gAudioFrames = context->audioFrames;
  gAudioInChannels = context->audioInChannels;
  
  gPipe.setup("sndfile-write", 65536, false, false);
  openFile();
  
  if((gFillBufferTask = Bela_createAuxiliaryTask(&writeBuffer, 90, "writeBuffer")) 
  == 0) {
    return false;
  }
  return true;
  
}
\end{lstlisting}

The setup function is a user-defined initialisation function which runs before audio rendering begins. This function runs once at the beginning of the program, after most of the system initialisation has begun but before audio rendering starts. This is used to prepare any memory or resources that will be needed in render. The audio sample rate, frames and audio input channels are all setup using the Bela context structure. The pipe is setup to write data with a specified size and whether the reads at the realtime and non-realtime side should be blocking. The openFile function is called to open the file for writing the audio data. The auxiliary task is then created with write buffer parameters. 

\vspace{0.2in}
\begin{lstlisting}[frame=single, language=C++]
void render(BelaContext* context, void* arg)
{
  gPipe.writeRt(context->audioIn, context->audioFrames * context->audioInChannels);
  Bela_scheduleAuxiliaryTask(gFillBufferTask);
}
\end{lstlisting}

The render function is a user-defined callback function to process audio and sensor data. This function is called regularly by the system every time there is a new block of audio and/or sensor data to process. The writeRt function reads data from the non-realtime side. The \lstinline{context->audioIn} is the float* that points to all the input samples, stored as interleaved channels. The audioIn is an array of 4 frames * 2 channels = 8 audio input samples. The auxiliary task which has previously been created is scheduled to run.

\vspace{0.2in}
\begin{lstlisting}[frame=single, language=C++]
void cleanup(BelaContext* context, void* arg)
{
    closeFile();
    free(uniqueFilename);
}
\end{lstlisting}

The cleanup function runs when the program finishes to free up memory. This function is called by the system once after audio rendering has finished, before the program quits. It is used to release any memory allocated in setup and to perform any other required cleanup. If no initialisation is performed in setup, then this function will usually be empty. The file is closed and the file that has been created by the generateUniqueFilename has its block of memory deallocated.

\bibliography{Ref}

\balance

\end{document}